\documentclass[sigconf]{acmart}
\usepackage{subfloat}
\usepackage{subfig}

\setcopyright{none}
\renewcommand\footnotetextcopyrightpermission[1]{} % removes footnote with conference information in first column
\acmConference[HSCC'20]{}{April 21-24, 2020}{Sydney, Australia}

\begin{document}

\title{Case Study: Verifying the Safety of an Autonomous Racing Car
  with a Neural Network Controller}

\author{Radoslav Ivanov, Taylor J. Carpenter, James Weimer, Rajeev Alur, George J. Pappas, Insup Lee}
\affiliation{%
  \institution{University of Pennsylvania}
  \streetaddress{3330 Walnut Street}
  \city{Philadelphia}
  \state{Pennsylvania}
}
\email{{rivanov, carptj, weimerj, alur, pappasg, lee}@seas.upenn.edu}

\renewcommand{\shortauthors}{R. Ivanov et al.}

%% This command allows the author to define a more concise list of
%% authors' names for this purpose.

\begin{abstract}
This paper describes a verification case study on an autonomous racing
car with a neural network (NN) controller. Although several
verification approaches have been proposed over the last year, they
have only been evaluated on low-dimensional systems or systems with
constrained environments. To explore the limits of existing
approaches, we present a challenging benchmark in which the NN takes
raw LiDAR measurements as input and outputs steering for the car. We
train a dozen NNs using two reinforcement learning algorithms and show
that the state of the art in verification can handle systems with
around 40 LiDAR rays, well short of a typical LiDAR scan with 1081
rays. Furthermore, we perform real experiments to investigate the
benefits and limitations of verification with respect to the sim2real
gap, i.e., the difference between a system's modeled and real
performance. We identify cases, similar to the modeled environment, in
which verification is strongly correlated with safe behavior. Finally,
we illustrate LiDAR fault patterns that can be used to develop robust
and safe reinforcement learning algorithms.
\end{abstract}

%\keywords{datasets, neural networks, gaze detection, text tagging}

\maketitle

\section{Introduction}
\label{sec:intro}
Neural networks (NNs) have shown great promise in multiple application
domains, including safety-critical systems such as autonomous
driving~\cite{bojarski16} and air traffic collision avoidance
systems~\cite{julian16}. At the same time, widespread adoption of
NN-based autonomous systems is hindered by the fact that NNs often
fail in seemingly unpredictable ways: slight perturbations in their
inputs can result in drastically different outputs, as is the case
with adversarial examples~\cite{szegedy13}. Such issues might lead to
fatal outcomes in safety-critical systems~\cite{uber_report} and thus
underscore the need to assure the safety of NN-based systems before
they can be deployed at scale.

One way to reason about the safety of such systems is to formally
verify safety properties of a NN's outputs for certain sensitive
inputs, as proposed in several NN verification and robustness
works~\cite{dutta18,ehlers17,katz17,wang18,weng18}. However, safety of
the NN does not immediately imply safety of the entire autonomous
system. A more exhaustive approach is to consider the interaction
between the NN and the physical plant (e.g., a car), trace the
evolution of the plant's states (e.g., position, velocity) over time
and ensure all reachable states are safe. A few such methods were
recently developed to verify safety of autonomous systems with NN
controllers~\cite{dutta19,ivanov19,sun19,tran19}. These techniques
combine ideas from classical dynamical system
reachability~\cite{chen13,kong15,tran19} (e.g., view the NN as a
hybrid system) with NN verification approaches (e.g., cast NN
verification as a mixed integer linear program). However, these
approaches have so far been evaluated on fairly simple systems: either
systems with low-dimensional NN inputs (the inputs are the plant
states, e.g., position and velocity~\cite{dutta19,ivanov19,tran19}) or
with constrained environments (e.g., LiDAR orientation does not change
over time~\cite{sun19}).

Two main challenges remain in applying verification techniques to
realistic systems. The first one is scalability. There are (at least)
two aspects to this challenge: 1) scalability with respect to (w.r.t)
the plant dynamics and 2) and scalability w.r.t. the NN
complexity. The reachability problem is undecidable for general hybrid
systems~\cite{alur95}, which means existing approaches can only
approximate the reachable sets. The NN adds additional complexity both
due to size and due to the number of inputs to the NN -- it is much
more challenging to compute reachable sets for multivariate functions,
even for small NNs. At the same time, having the capability to verify
systems with high-dimensional measurements is crucial, since NNs are
most useful exactly in such settings.

The second verification challenge is the sim2real gap, i.e., the
difference between a system's modeled and real
performance~\cite{chebotar19}. Analyzing the sim2real gap is essential
as it allows us to explore the benefit of verification with the
respect to the real system. Overcoming this challenge would enable
developers to design and test approaches in simulation with the
assurance that safety properties that hold in simulation would carry
over to the real world.

In order to illustrate these difficulties and to provide a challenging
benchmark for future work, this paper presents a verification case
study on a realistic NN-controlled autonomous system. In particular,
we focus on the F1/10 autonomous racing car~\cite{f1tenth}, which
needs to navigate a structured environment using high-dimensional
LiDAR measurements. This case study has two goals: 1) assess the
capabilities of existing verification approaches and highlight aspects
that require future work; 2) investigate conditions under which the
verification translates to safe performance in the real world.

To perform the verification, we first identify a dynamics (bicycle)
model of the F1/10 car, as well as an observation model mapping the
vehicle state to the LiDAR measurements. To obtain the observation
model, we assume the car operates in a structured environment (i.e., a
sequence of hallways) such that each LiDAR ray can be calculated based
on the car's state and the surrounding walls. Given these models, we
train an end-to-end NN controller using reinforcement
learning~\cite{lillicrap15}. The controller takes LiDAR measurements
as input and produces steering commands as output (assuming constant
throttle). Once the NN is trained, we aim to verify that the car does
not crash in the hallway walls.

We evaluate the scalability of existing verification approaches by
varying the NN size, the number of LiDAR rays as well as the training
algorithm. Note that the complexity of the verification task grows
exponentially with the number of rays since, depending on the
uncertainty, a given ray could reach different walls, which triggers
multiple paths in the hybrid observation model that need to be
verified simultaneously. We use the state-of-the-art tool
Verisig~\cite{ivanov19} to verify the dozen setups that were trained;
we could not encode the hybrid LiDAR model in the other existing
tools. In our evaluation, Verisig could handle NNs of roughly 250
neurons (containing two layers with 128 neurons each) and LiDAR scans
with around 40 rays. This highlights the challenge of this
verification task: verifying the entire LiDAR scan containing 1081
rays, together with a corresponding NN that can effectively process
such a scan, remains well beyond the capabilities of existing tools.

Finally, we perform experiments, using the verified controllers, to
evaluate the system's sim2real gap. This gap is especially pronounced
with LiDAR measurements, since rays can get reflected depending on the
reached surface, thereby providing an erroneous distance. We first
evaluate the benefit of verification in an ideal setting by performing
experiments with all reflective surfaces covered -- all NNs performed
similarly in this setup, resulting in safe behavior roughly 90\% of
the time; even in these cases crashes were caused by LiDAR faults that
could not be completely eliminated. However, more crashes were
observed in the unmodified environment, as caused by consistently bad
LiDAR data. Upon closer investigation, we identified patterns of LiDAR
faults that reproduce the unsafe behavior in simulations as well --
however, we could not train a robust controller using the
state-of-the-art reinforcement learning algorithms. Thus, it remains
an open problem to train (and verify) a NN that provides safe
performance in the presence of LiDAR errors.

In summary, this paper has three contributions: 1) a challenging
benchmark for verification and reinforcement learning in
neural-network-controlled autonomous systems with high-dimensional
measurements; 2) an exhaustive evaluation of a state-of-the-art
verification tool on this benchmark; 3) real experiments that
illustrate the benefits and limitations of verification w.r.t. the
sim2real gap.

\begin{figure*}[t]
  \begin{minipage}[t]{0.33\textwidth}
  \centering
  \includegraphics[width=\linewidth]{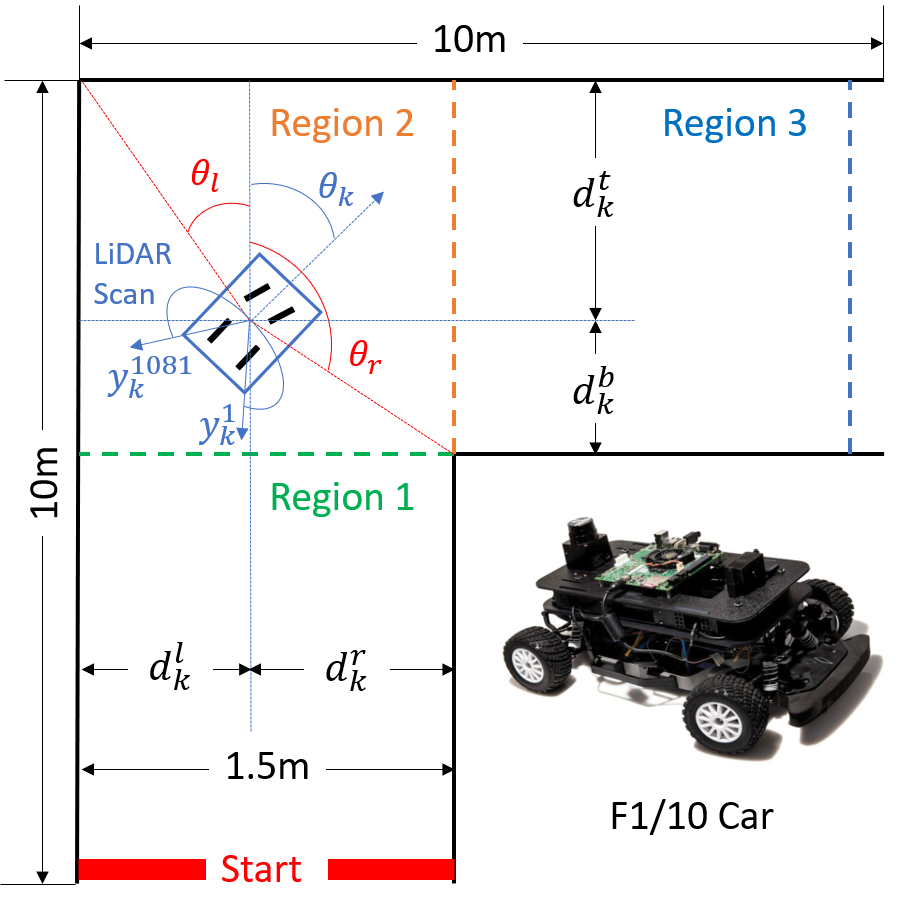}
  \caption{Navigation scenario. There are three different regions
    depending on how many walls can be reached using LiDAR.}
  \label{fig:hallways}
\end{minipage}%
\hfill%
\begin{minipage}[t]{0.6\textwidth}
  \centering 
  \includegraphics[width=\linewidth]{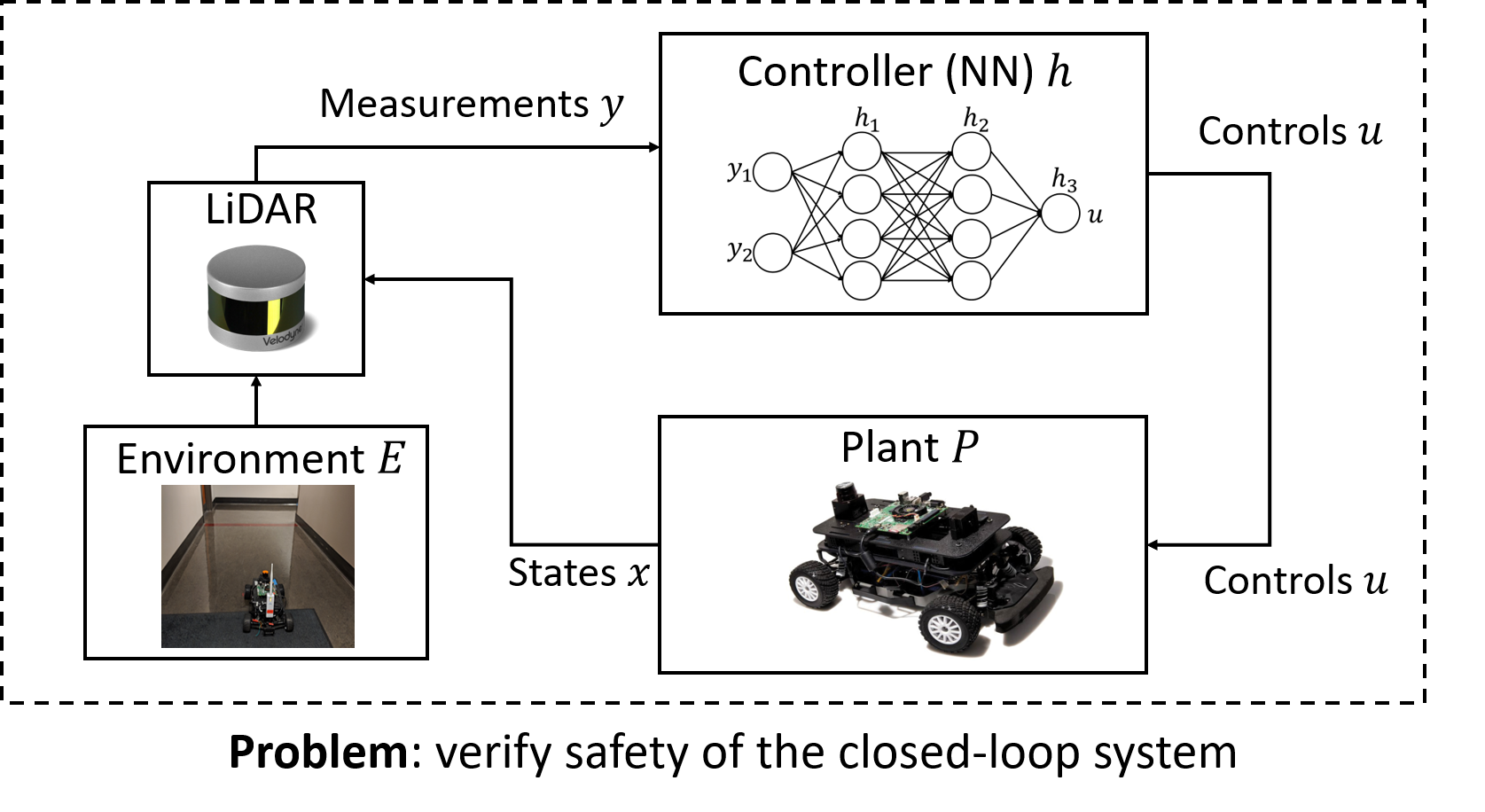}
  \caption{Overview of the closed-loop system and the problem
    considered in this paper.}
  \label{fig:problem}
\end{minipage}
\end{figure*}

\section{System Overview}
\label{sec:background}
This section summarizes the different parts of the case study
considered in the paper. We first describe the F1/10 platform,
followed by a high-level introduction to reinforcement learning and
hybrid system verification.

\subsection{The F1/10 Autonomous Racing Car}
The case study considered in this paper is inspired by the F1/10
Autonomous Racing Competition~\cite{f1tenth}, where an autonomous car
must navigate a structured environment (i.e., the track) as fast as
possible. The F1/10 car is shown in Figure~\ref{fig:hallways}. It is
built for racing purposes and can reach up to 40mph. The car is
controlled by an onboard chip such as the NVIDIA Jetson TX2 module.

A diagram of the closed-loop system is shown in
Figure~\ref{fig:problem}. The car operates in a hallway environment;
without loss of generality, we assume that all turns are 90-degree
right turns such that the ``track'' is effectively a square. Although
in the competition the car has access to a number of sensors, in this
case study we assume the controller only has access to LiDAR
measurements. The measurements are sent to a NN controller that
outputs a steering command to the vehicle. We assume that the car
operates at constant throttle, in order to keep the dynamics model and
the verification task manageable. The car's dynamic and observation
models used in the case study are described in
Section~\ref{sec:model}.

\subsection{Reinforcement Learning}
Overall, developing a robust controller for the F1/10 car is a
challenging task, both due to the difficulty of analyzing LiDAR
measurements and to the speed and agility of the car. Thus, this is a
good application for reinforcement learning~\cite{lillicrap15}, where
no knowledge of the car dynamics or the observation model is
required. In reinforcement learning, the controller starts by applying
a control action and observing a reward. As training proceeds, the
learning problem is to maximize the reward by exploring the state
space and trying different controls. In recent years, deep
reinforcement learning (where controllers are neural networks) has
shown great promise in a number of traditionally challenging problems,
such as playing Atari games~\cite{mnih15}, controlling autonomous
cars~\cite{bojarski16} and playing board games~\cite{silver16}. Hence,
reinforcement learning is a natural choice for learning a controller
for the F1/10 car as well; the specific training approach is described
in Section~\ref{sec:training}.

\subsection{Hybrid System and NN Verification}
The hybrid system verification problem can be stated at a high level
as follows: given a hybrid model of the plant dynamics and
observations, the problem is to trace the evolution of the plant
states over time (for a set of initial conditions) and verify that no
unsafe states can be reached. Although hybrid system reachability is
undecidable except for special cases such as linear
systems~\cite{alur95,lafferriere99} (see~\cite{alur11,doyen18} for an
exhaustive discussion), several approaches work well for specific
non-linear systems. In particular, reachability is $\delta$-decidable
for Type 2 computable functions~\cite{kong15}, which has led to the
development of the tool dReach. Alternatively, Flow*~\cite{chen13}
constructs Taylor model (TM) approximations of the system's reachable
set. While Flow* provides no decidability claims, it can verify
interesting properties for multiple non-linear systems classes and
scales well when using TMs with interval analysis.

Recently, several approaches were developed for verification of hybrid
systems with NNs controllers~\cite{dutta19,ivanov19,sun19,tran19}. As
described in Section~\ref{sec:intro}, the NN introduces new challenges
both due to its size and complexity. To address this issue, the
proposed approaches borrow ideas from classical hybrid system
reachability, e.g., by transforming the NN into a mixed-integer linear
program (MILP)~\cite{dutta19}, a satisfiability modulo theory (SMT)
formula~\cite{sun19} or an equivalent hybrid
system~\cite{ivanov19}. Although existing tools have shown promising
scalability in terms of the size of the NN, they have only been
evaluated on low-dimensional systems or systems with constrained
environments. In this paper, we provide a much more challenging
scenario, with a high-dimensional hybrid observation model, in order
to test the limits of these tools and to highlight avenues for future
work.

\subsection{System Design and Development}
In order to build and verify the system, we perform the following
steps: 1) build a model of the car dynamics and observations; 2) train
a NN on the model using reinforcement learning; 3) verify that the
NN-controlled car is safe with respect to the model; 4) perform real
experiments to analyze the sim2real gap. The following sections
describe each of these steps in more detail.

\section{Plant Model}
\label{sec:model}
This section describes the F1/10 car's dynamical and observation
models. These models are used to train the NN controller
(Section~\ref{sec:training}) and to perform the closed-loop system
verification (Section~\ref{sec:verification}).

\subsection{Dynamics model}
We use a bicycle model~\cite{polack17,rajamani11} to model the car's
dynamics, which is a standard model for cars with front
steering. Specifically, we use a kinematic bicycle model since it has
few parameters (that are easy to identify) and tracks reasonably well
at low speeds, i.e., under 5 m/s~\cite{rajamani11}. In the kinematic
bicycle model, the car has four states: position in two dimensions,
linear velocity and heading. The continuous-time dynamics are given by
the following equations:
\begin{align}
  \begin{split}
    \dot{x} &= v cos(\theta + \beta)\\
    \dot{y} &= v sin(\theta + \beta)\\
    \dot{v} &= - c_av + c_ac_m(u - c_h)\\
    \dot{\theta} &= \frac{V cos(\beta)}{l_f + l_r}tan(\delta)\\
    \beta &= tan^{-1}\left(\frac{l_r tan(\delta)}{l_f + l_r}\right),\\
  \end{split}
\end{align}
where $v$ is the car's linear velocity, $\theta$ is the car's
orientation, $\beta$ is the car's slip angle and $x$ and $y$ are the
car's position; $u$ is the throttle input, and $\delta$ is the heading
input; $c_a$ is an acceleration constant, $c_m$ is a car motor
constant, $c_h$ is a hysteresis constant, and $l_f$ and $l_r$ are the
distances from the car's center of mass to the front and rear,
respectively. Since $tan^{-1}$ is not supported by most hybrid system
verification tools, we assume that $\beta = 0$; this is not a limiting
assumption in the considered case study as the slip angle is typically
fairly small at low speeds; we did not observe significant differences
in the model's predictive power due to this assumption. After
performing system identification, we obtained the following parameter
values: ${c_a = 1.633}, {c_m = 0.2}, {c_h = 4}, {l_f = 0.225m}, {l_r =
  0.225m}$. Finally, we assume a constant throttle $u = 16$ (resulting
in a top speed of roughly 2.4 m/s), i.e., the controller only controls
heading. We emphasize that the plant model is fairly non-linear, thus
making it difficult to compute reachable sets for the car's states.

\subsection{Observation model}
The F1/10 car has access to LiDAR measurements only. As shown in
Figure~\ref{fig:hallways}, a typical LiDAR scan consists of a number
of rays emanating from -135 to 135 degrees relative to the car's
heading. For each ray, the car receives the distance to the first
obstacle the ray hits; if there are no obstacles within the LiDAR
range, the car receives the maximum range. In this case study, we
consider a LiDAR scan with a maximum of 1081 rays and a range of 5
meters.\footnote{Although typical LiDARs have a longer range than 5m,
  we found our unit's measurements to be unreliable beyond 5m.}

As shown in Figure~\ref{fig:hallways}, there are three possible
regions the car can be in, depending on how many walls can be reached
using LiDAR. The worst case is Region 2, in which there are four walls
to consider. We present the measurement model for Region 2 only since
the other regions are special cases of Region 2. Let $\alpha_1, \dots,
\alpha_{1081}$ denote the relative angles for each ray with respect to
the car's heading, i.e., $\alpha_1 = -135, \alpha_2 = -134.75, \dots,
\alpha_{1081} = 135$. One can determine which wall each LiDAR ray hits
by comparing the $\alpha_i$ for that ray with the relative angles to
the two corners of that turn, $\theta_l$ and $\theta_r$ in
Figure~\ref{fig:hallways}. The measurement model for Region 2 (for a
right turn) is presented below, for $i \in \{1, \dots, 1081\}$:
\begin{align}
  \begin{split}
    y_k^i = \left\{ \begin{array}{lll} d^r_k/cos(90 + \theta_k + \alpha_i) &&\text{if } \theta_k + \alpha_i \le \theta_r \\
      d^b_k/cos(180 + \theta_k + \alpha_i) &&\text{if } \theta_r < \theta_k + \alpha_i \le -90 \\
      d^t_k/cos(\theta_k + \alpha_i) &&\text{if } -90 < \theta_k + \alpha_i \le \theta_l \\
      d^l_k/cos(90 - \theta_k - \alpha_i) &&\text{if } \theta_l < \theta_k + \alpha_i,\end{array} \right.
  \end{split}
\end{align}
where $k$ is the sampling step (the sampling rate is assumed to be
10Hz), $d^t_k, d^b_k, d^l_k, d^r_k$ are distances to the four walls,
as illustrated in Figure~\ref{fig:hallways}, and can be derived from
the car's position $(x,y)$.\footnote{If $\theta_k + \alpha_i \notin
  [-180,180]$, $\theta_k + \alpha_i$ needs to be normalized by
  adding/subtracting 360.} Similar to the dynamics model, the
measurement model is non-linear. Furthermore, keeping the
approximation error small during the reachability analysis is
challenging since if a ray is almost parallel to a wall, small
uncertainty in the car's heading results in large uncertainty in the
measured distance for that ray, which is evident in the division by
cosine in the measurement model.

\begin{table*}
\centering
\begin{tabular}{|l|c|c|c|c|c|c|c|}
\hline
DRL algorithm & NN setup & \# LiDAR rays & Controller index  & Initial interval size  & NN ver. time (s) & Total ver. time (s) & \# paths \\ \hline
DDPG & $64 \times 64$ & 21 & 1 & 0.2cm & 355  & 4126  & 1.32  \\ \hline
DDPG & $64 \times 64$ & 21 & 2 & 0.2cm & 347  & 4122  & 1.15  \\ \hline
DDPG & $64 \times 64$ & 21 & 3 & \multicolumn{4}{c|}{DNF}  \\ \hline
DDPG & $128 \times 128$ & 21 & 1 & \multicolumn{4}{c|}{DNF}  \\ \hline
DDPG & $128 \times 128$ & 21 & 2 & \multicolumn{4}{c|}{DNF}  \\ \hline
DDPG & $128 \times 128$ & 21 & 3 & \multicolumn{4}{c|}{DNF}  \\ \hline
TD3 & $64 \times 64$ & 21 & 1 & 0.5cm & 553  & 4731  & 2.2  \\ \hline
TD3 & $64 \times 64$ & 21 & 2 & 0.5cm & 853  & 8094  & 2.825  \\ \hline
TD3 & $64 \times 64$ & 21 & 3 & 0.5cm & 724  & 8641  & 2.725  \\ \hline
TD3 & $128 \times 128$ & 21 & 1 & 0.5cm & 197  & 3760  & 1.6  \\ \hline
TD3 & $128 \times 128$ & 21 & 2 & 0.5cm & 355  & 5954  & 1.775  \\ \hline
TD3 & $128 \times 128$ & 21 & 3 & \multicolumn{4}{c|}{Verisig/Flow* crash}  \\ \hline
TD3 & $64 \times 64$ & 41 & 1 & 0.2cm* & 634  & 11915  & 2.194  \\ \hline
TD3 & $128 \times 128$ & 41 & 1 & \multicolumn{4}{c|}{DNF}  \\ \hline
TD3 & $64 \times 64$ & 61 & 1 & \multicolumn{4}{c|}{DNF}  \\ \hline
TD3 & $128 \times 128$ & 61 & 1 & \multicolumn{4}{c|}{DNF}  \\ \hline
\end{tabular}
\caption{Verification evaluation for various setups in terms of NN
  architectures and number of LiDAR rays. The verification times and
  the number of paths are averaged over all subsets for each
  setup. Subset sizes are decreased from 0.5cm to 0.2cm and to 0.1cm,
  if verification fails. DNF setups were terminated after 10 hours on
  0.1cm subsets. The notation $n \times n$ means that this is a NN
  with two hidden layers and $n$ neurons per layer. Note that 2 out of
  100 instances of the 41-ray setup did not terminate within 24 hours
  (and were killed), which highlights the complexity of the larger
  setup.}
\label{tab:verification}
\vspace{-20pt}
\end{table*}

\section{Controller Training}
\label{sec:training}
As mentioned in Section~\ref{sec:background}, the F1/10 case study is
a good application domain for deep reinforcement learning (DRL) due to
the high-dimensional measurements as well as the non-trivial control
policy that is required. This section discusses the DRL algorithms
used in the case study as well as the choice of reward function.

Multiple DRL algorithms have been proposed in the past few years,
depending on the learning setup. In settings with a discrete number of
control actions, the standard approach is to use a deep
Q-network~\cite{mnih15}, as inspired by the idea of Q learning, i.e.,
learning the (Q) function that maps a state and an action to the
maximum expected reward that can be achieved by taking that action. In
the case of continuous actions, a deep deterministic policy gradient
(DDPG) approach~\cite{lillicrap15} was developed that approximates the
Q function using a Bellman equation. Notably, DDPG uses two NNs while
training, a critic that learns the Q function and an actor that
applies the controls. Once training is finished, only the actor is
used as the actual controller. Multiple approaches have been proposed
to improve upon DDPG, especially in terms of training stability, e.g.,
using normalized advanced functions (NAFs)~\cite{gu16}, which are a
continuous version of Q functions, or using a twin delayed deep
deterministic policy gradient (TD3) algorithm~\cite{fujimoto18} that
employs two critics for greater stability. Finally, model-based DRL
algorithms have also been proposed where the NN architecture is
designed to implicitly learn the plant model in order to improve
training~\cite{finn17}.

In this paper, we focus on the continuous-action-space algorithms as
they fit better the F1/10 car control task. For better evaluation, we
train controllers using two different algorithms, namely DDPG and TD3
(we could not train good controllers using the authors' implementation
of the NAF-based approach).

An important consideration in any DRL problem is the choice of reward
function. In particular, we are interested in a reward function that
not only results in better training but also in ``smooth'' control
policies that are easier to verify. Thus, the reward function consists
of two parts: 1) a positive gain for every step that does not result
in a crash (to enforce safe control) and 2) a negative gain penalizing
higher control inputs (to enforce smooth control):
\begin{equation}
r_k = g_p - g_n \delta_k^2,
\end{equation}
where $g_p = 10$, $g_n = 0.05$. A large negative reward of -100 is
received if the car crashes. Note that the negative input gain is not
applied in turns in order to avoid a local optimum while training.

Another hyper parameter in the training setup is the NN
architecture. Although convolutional NNs are easier to train with
high-dimensional inputs, they are harder to verify by existing tools
since each convolutional layer needs to be unrolled in a fully
connected layer with a large number of neurons. Thus, we only consider
fully connected architectures in this case study. Scaling to
convolutional NNs is thus an important avenue for future work in NN
verification.

\begin{figure*}[!t]
  \centering 
       \subfloat[DDPG, $64 \times 64$, controller 1.]{
           \includegraphics[width=0.24\linewidth]{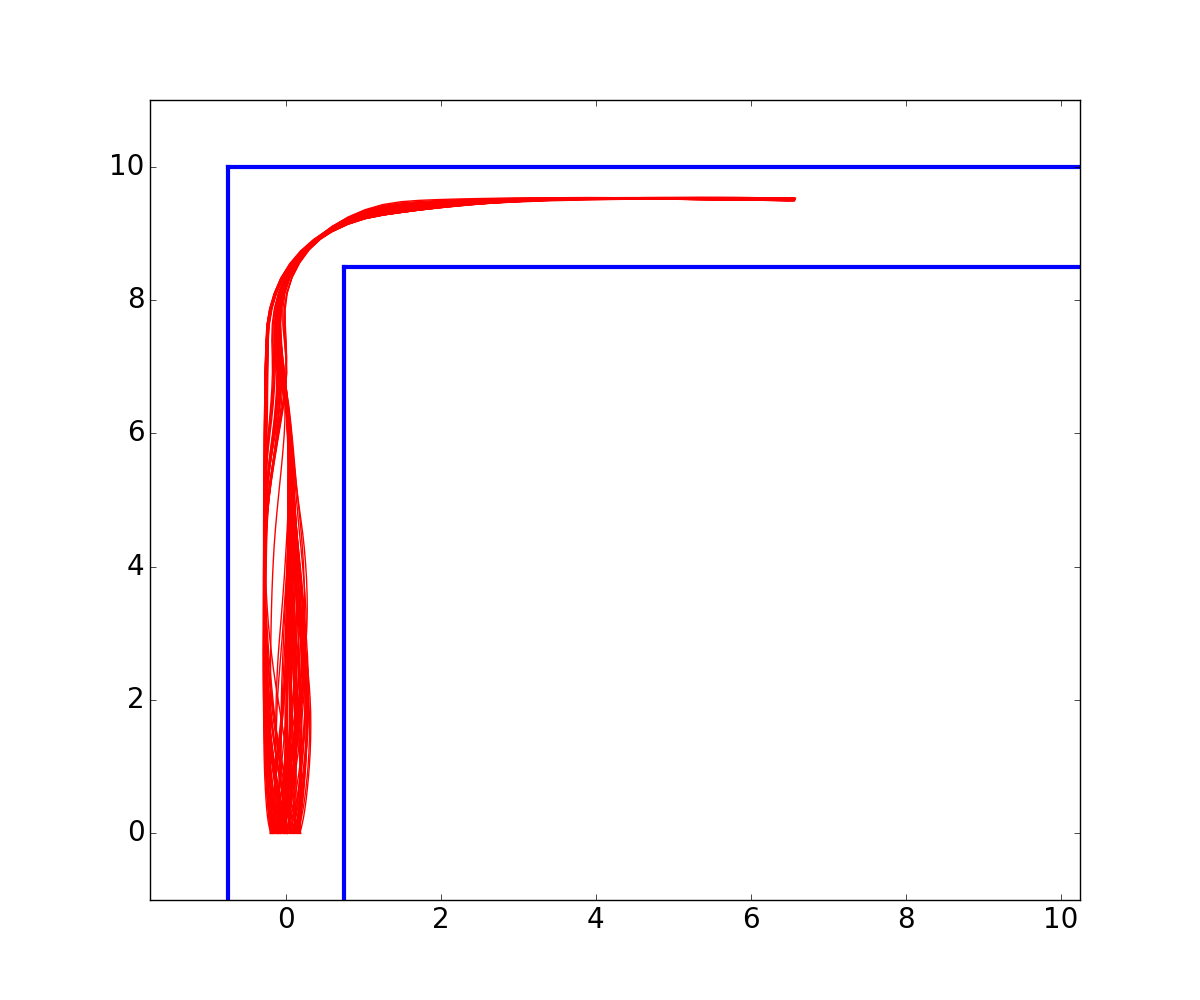}
           \label{fig:tanh64x64_DDPG1}
       }
       \subfloat[DDPG, $128 \times 128$, controller 2.]{
           \includegraphics[width=0.24\linewidth]{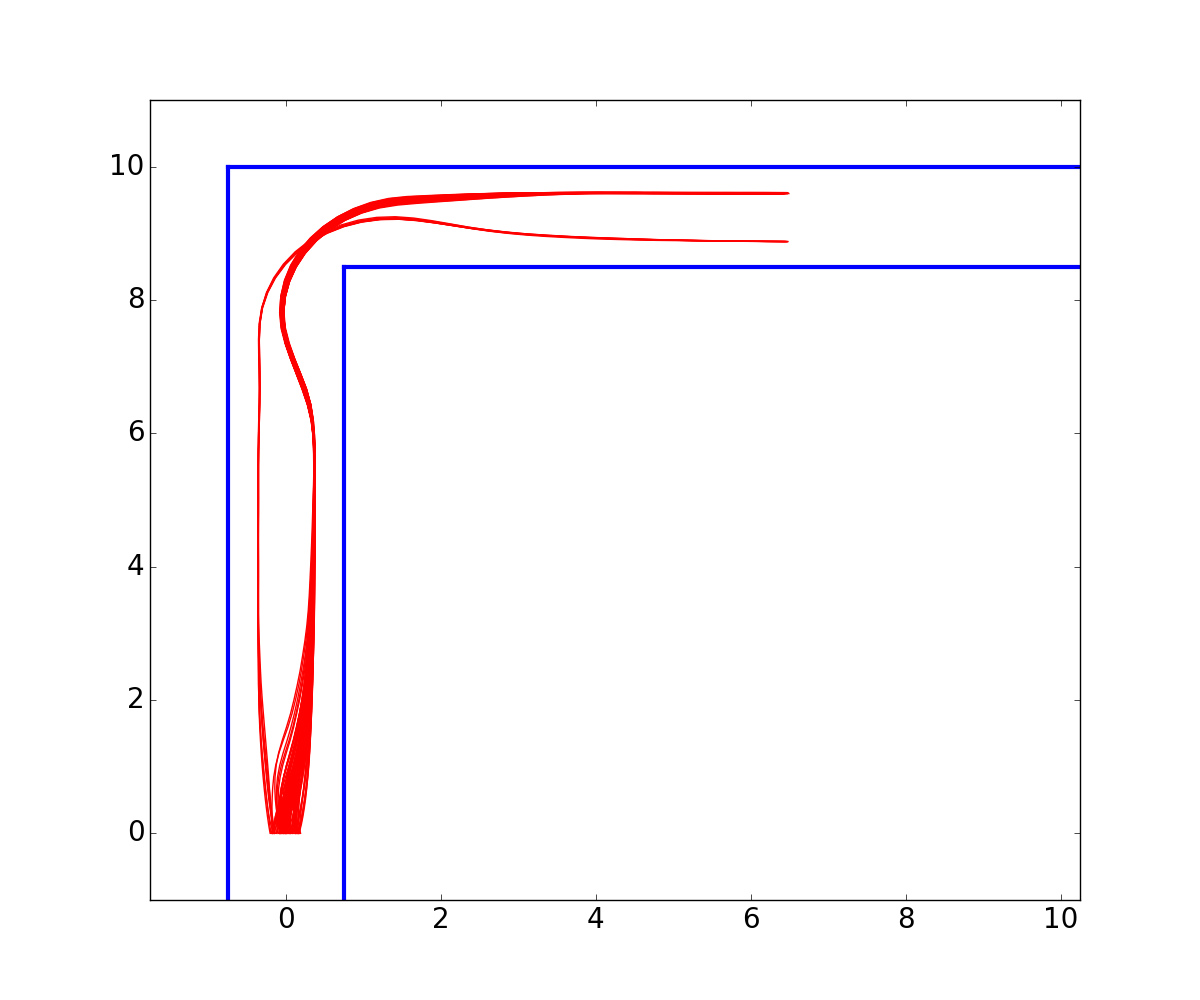}
           \label{fig:tanh128x128_DDPG2}
       }
       \subfloat[TD3, $64 \times 64$, controller 1.]{
           \includegraphics[width=0.24\linewidth]{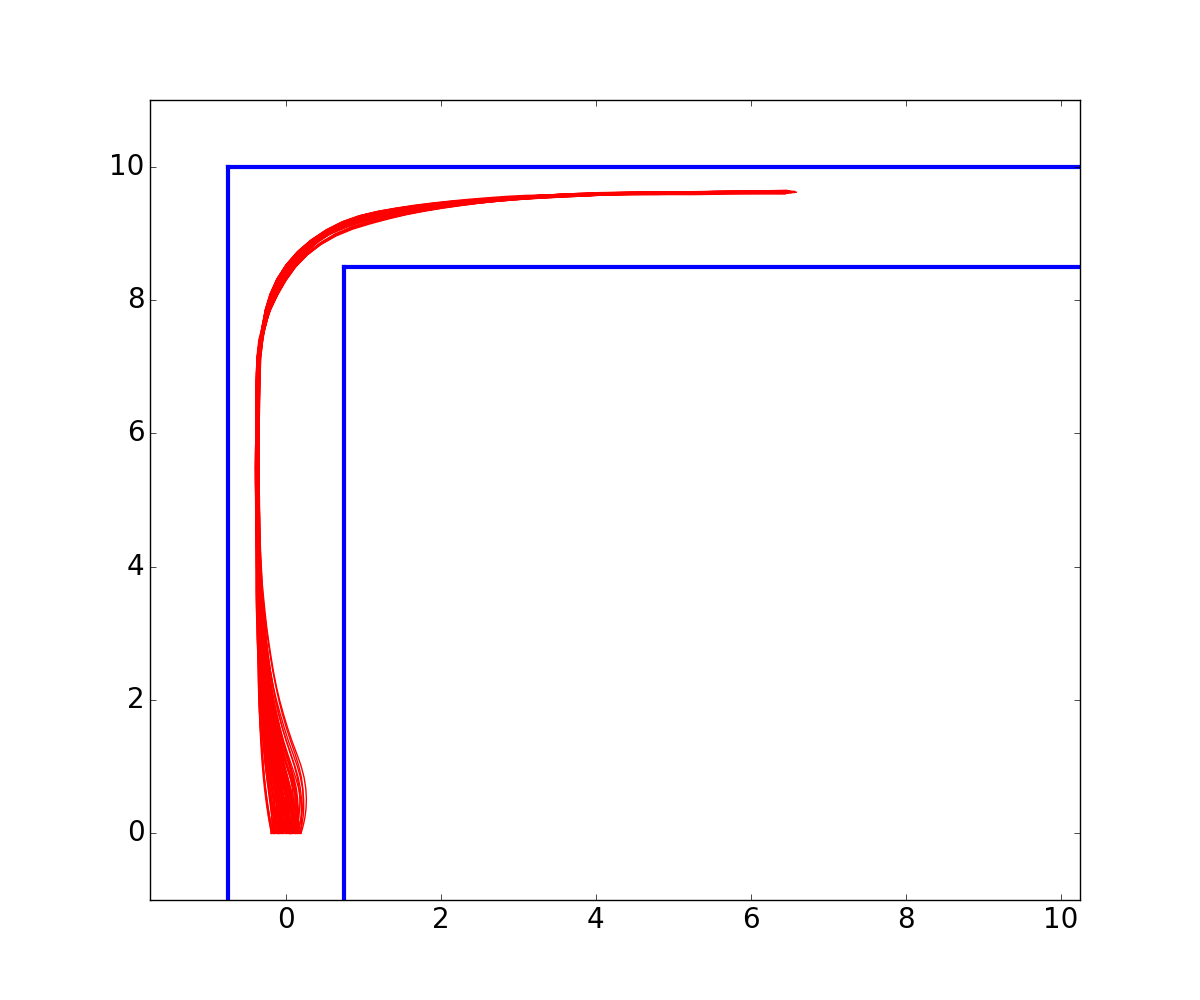}
           \label{fig:tanh64x64_TD3_1}
       }
       \subfloat[TD3, $128 \times 128$, controller 1.]{
           \includegraphics[width=0.24\linewidth]{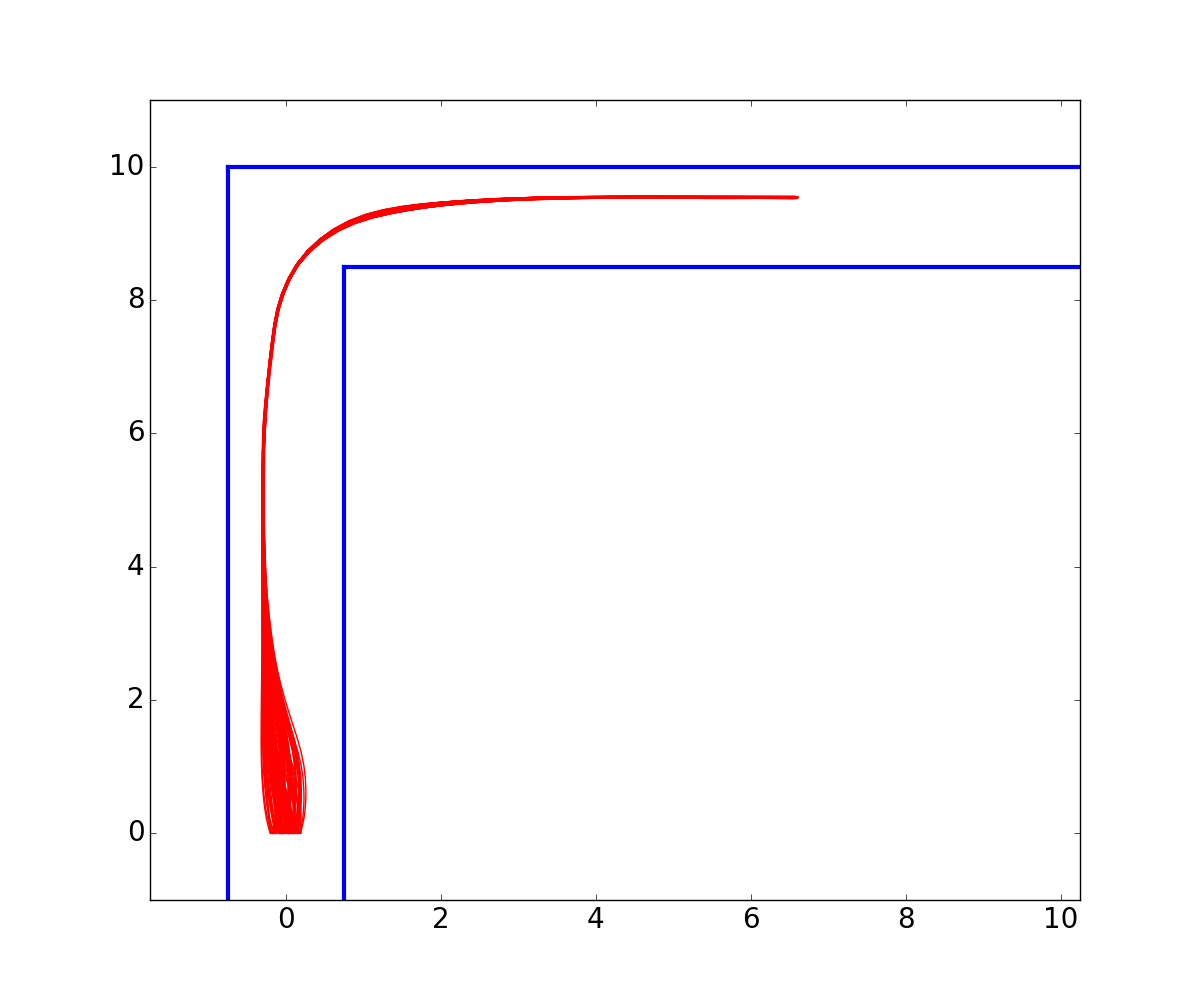}
           \label{fig:tanh128x128_TD3_1}
       }
       \vspace{-10pt}
       \caption{Simulation traces for different NN controllers from
         Table~\ref{tab:verification}.}
  \label{fig:sim_traces}
\vspace{-10pt}
\end{figure*}

\begin{figure}[!t]
  \centering 
       \subfloat[Modified environment.]{
           \includegraphics[width=0.45\linewidth]{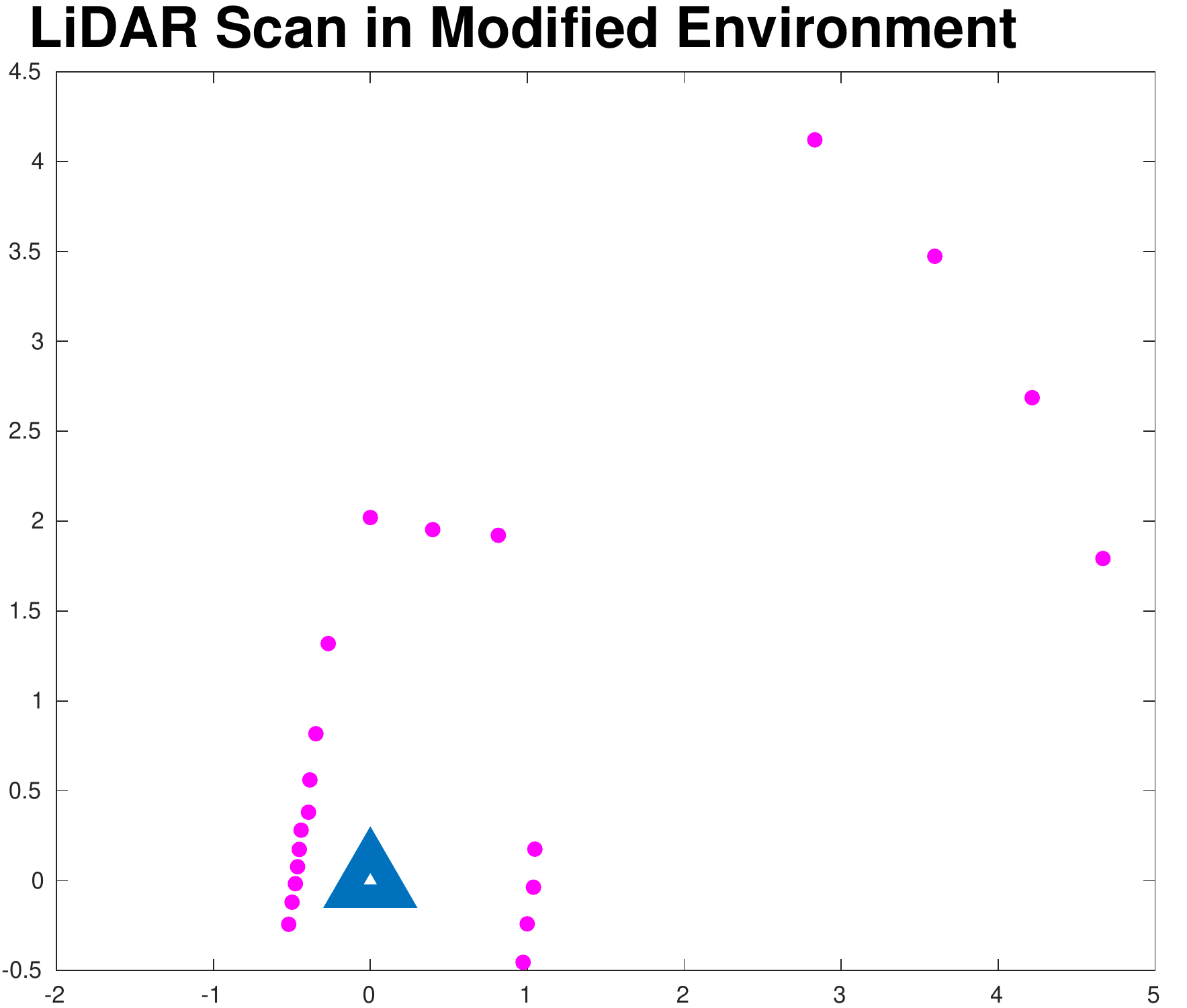}
           \label{fig:scan_modified_environment}
       }
       \hfill
       \hfill
       \subfloat[Unmodified environment.]{
           \includegraphics[width=0.47\linewidth]{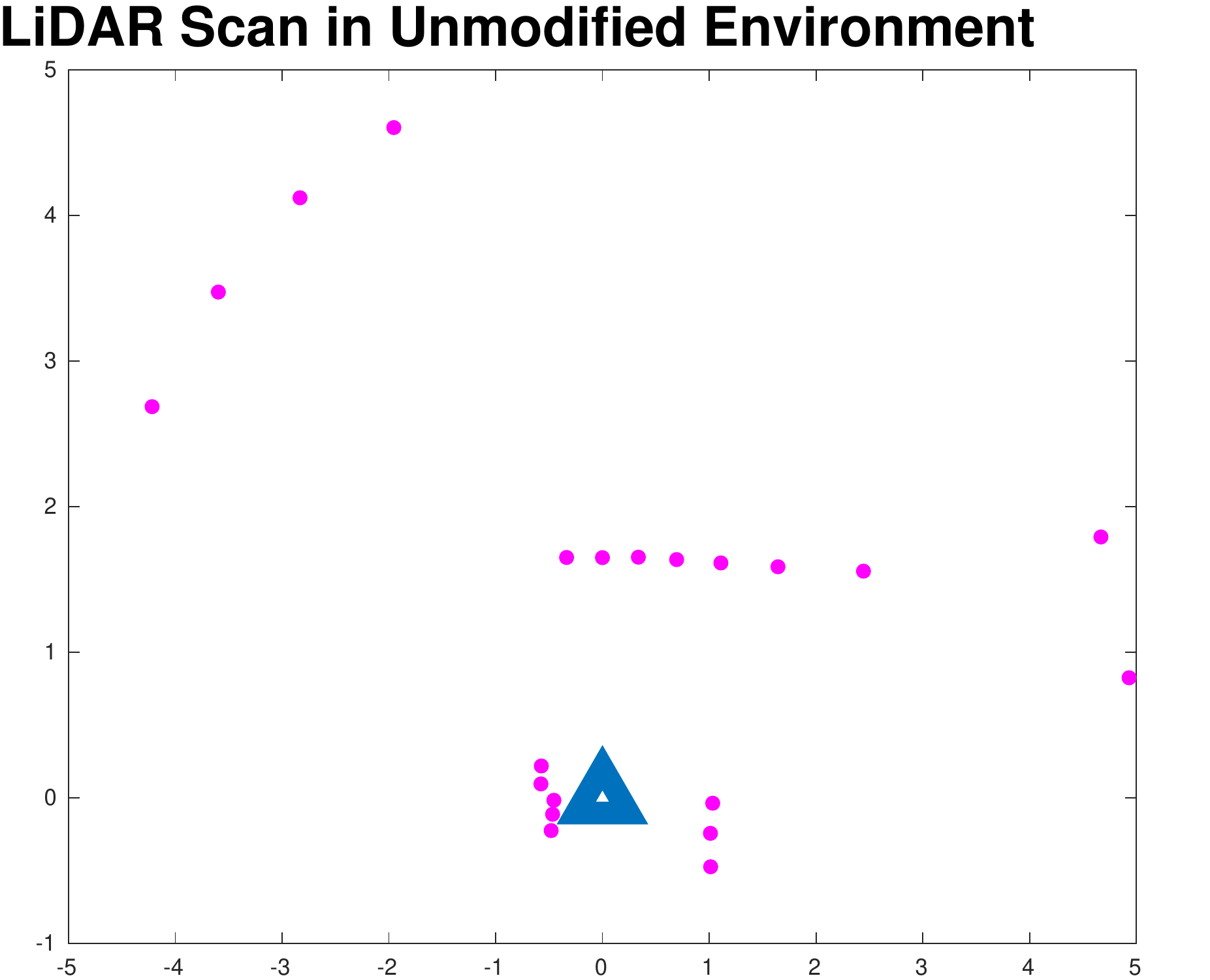}
           \label{fig:scan_unmodified_environment}
       }
       \caption{LiDAR scans that led to crashes in experiments.}
  \label{fig:lidar_scans}
\vspace{-10pt}
\end{figure}

\section{Verification Evaluation}
\label{sec:verification}
Having described the NN controller training process, we now evaluate
the scalability of a state-of-the-art verification tool,
Verisig~\cite{ivanov19}. As mentioned in Section~\ref{sec:intro}, the
other existing tools cannot currently handle the hybrid observation
model. In the considered scenario, the car starts from a 20cm-wide
range in the middle of the hallway (as illustrated in
Figure~\ref{fig:hallways}) and runs for 7s. This is enough time for
the car to reach top speed before the first turn and to get roughly to
the middle of the next hallway. The safety property to be verified is
that car is never within 0.3m of either wall.

Verisig can only handle NNs with smooth activation functions (i.e.,
sigmoid and hyperbolic tangent) and works by transforming the NN into
an equivalent hybrid system. The NN's hybrid system is then composed
with the plant's hybrid system, thereby casting the problem as a
hybrid system verification problem that is solved by Flow*. In
Verisig's original evaluation~\cite{ivanov19}, the tool scales to NNs
with about 100 neurons per layer and a dozen layers. The
high-dimensional input space considered in this case study, however,
presents a greater challenge which might also affect the tool's
scalability in terms of the NN size.

Since Verisig only accepts NNs with smooth activations, all NNs in
this case study have hyperbolic tangent (tanh) activations. The output
layer also has a tanh activation, which is scaled by 15 so that the
control input ranges from -15 to 15 degrees.\footnote{The dynamics
  model assumes the controls are given in radians -- we use degrees in
  the paper for clearer presentation.} As described in
Section~\ref{sec:training}, we use both the DDPG and TD3 algorithms to
explore different aspects of the verification process. All NNs have
two hidden fully connected layers; the number of neurons per layer is
increased from 64 to 128. We also vary the number of LiDAR rays from
21 to 41 and finally to 61 in order to evaluate the scalability in
terms of the input dimension as well.\footnote{Note that, due to
  hardware issues with our LiDAR unit, we only used the rays ranging
  from -115 to 115 degrees (instead of the full scan ranging from -135
  to 135 degrees).} For repeatability purposes, we train three
controllers for each setup in the 21-ray case.

The verification times\footnote{All experiments were run on a 80-core
  machine running at 1.2GHz. However, Flow* is not parallelized, so
  the only benefit from the multicore processor is the fact that
  multiple verification instances can be run at the same time.} for
all the setups are presented in Table~\ref{tab:verification}, together
with other verification artifacts. Note that the initial interval is
split in smaller subsets in order to maintain the approximation error
small -- the verification is performed separately for each subset. For
each setup, only average statistics over all subsets are presented. As
can be seen in the table, the biggest setup that Verisig can handle
has roughly 40 LiDAR rays. The verification complexity in terms of the
number of LiDAR rays is reflected in the number-of-paths column in the
table, which indicates the average number of paths in the hybrid
observation model caused by the fact that a LiDAR ray could
potentially reach different walls -- note that smaller-NN setups can
take longer to verify simply due to a higher number of paths since
each path needs to be verified separately.

\begin{table*}
\centering
\begin{tabular}{|l|c|c|c|c|c|}
\hline
DRL algorithm & NN architecture & \# LiDAR rays & Controller Index  & Safe outcomes in $Env_M$  & Safe outcomes in $Env_U$ \\ \hline
DDPG & $64 \times 64$ & 21 & 1 & 9/10 & 0/10    \\ \hline
DDPG & $64 \times 64$ & 21 & 2 & 9/10 & 2/10   \\ \hline
DDPG & $64 \times 64$ & 21 & 3 & 10/10 & 8/10   \\ \hline
DDPG & $128 \times 128$ & 21 & 1 & 10/10 & 8/10  \\ \hline
DDPG & $128 \times 128$ & 21 & 2 & 7/10 & 4/10  \\ \hline
DDPG & $128 \times 128$ & 21 & 3 & 9/10 & 0/10  \\ \hline
TD3 & $64 \times 64$ & 21 & 1 & 8/10 & 9/10  \\ \hline
TD3 & $64 \times 64$ & 21 & 2 & 10/10 & 9/10 \\ \hline
TD3 & $64 \times 64$ & 21 & 3 & 10/10 & 9/10   \\ \hline
TD3 & $128 \times 128$ & 21 & 1 & 9/10 & 9/10  \\ \hline
TD3 & $128 \times 128$ & 21 & 2 & 9/10 & 9/10   \\ \hline
TD3 & $128 \times 128$ & 21 & 3 & 9/10 & 9/10   \\ \hline
%% TD3 & $64 \times 64$ & 41 & 1 & 10/10 & 1/10  \\ \hline
%% TD3 & $128 \times 128$ & 41 & 1 & 10/10 & 10/10  \\ \hline
%% TD3 & $64 \times 64$ & 61 & 1 & 10/10 & 7/10    \\ \hline
%% TD3 & $128 \times 128$ & 61 & 1 & 10/10 & 10/10  \\ \hline
\end{tabular}
\caption{Sim2real gap for the 21-ray setups from
  Table~\ref{tab:verification}. Ten 7s runs were performed for each
  setup in each environment, where $Env_M$ and $Env_U$ refer to the
  modified and unmodified environments, respectively. A safe outcome
  is recorded if the car does not hit a wall during a run.}
\label{tab:sim2real}
\vspace{-35pt}
\end{table*}

\begin{figure*}[!th]
  \centering 
       \subfloat[DDPG, $64 \times 64$, controller 1: 24 \% safe.]{
           \includegraphics[width=0.23\linewidth]{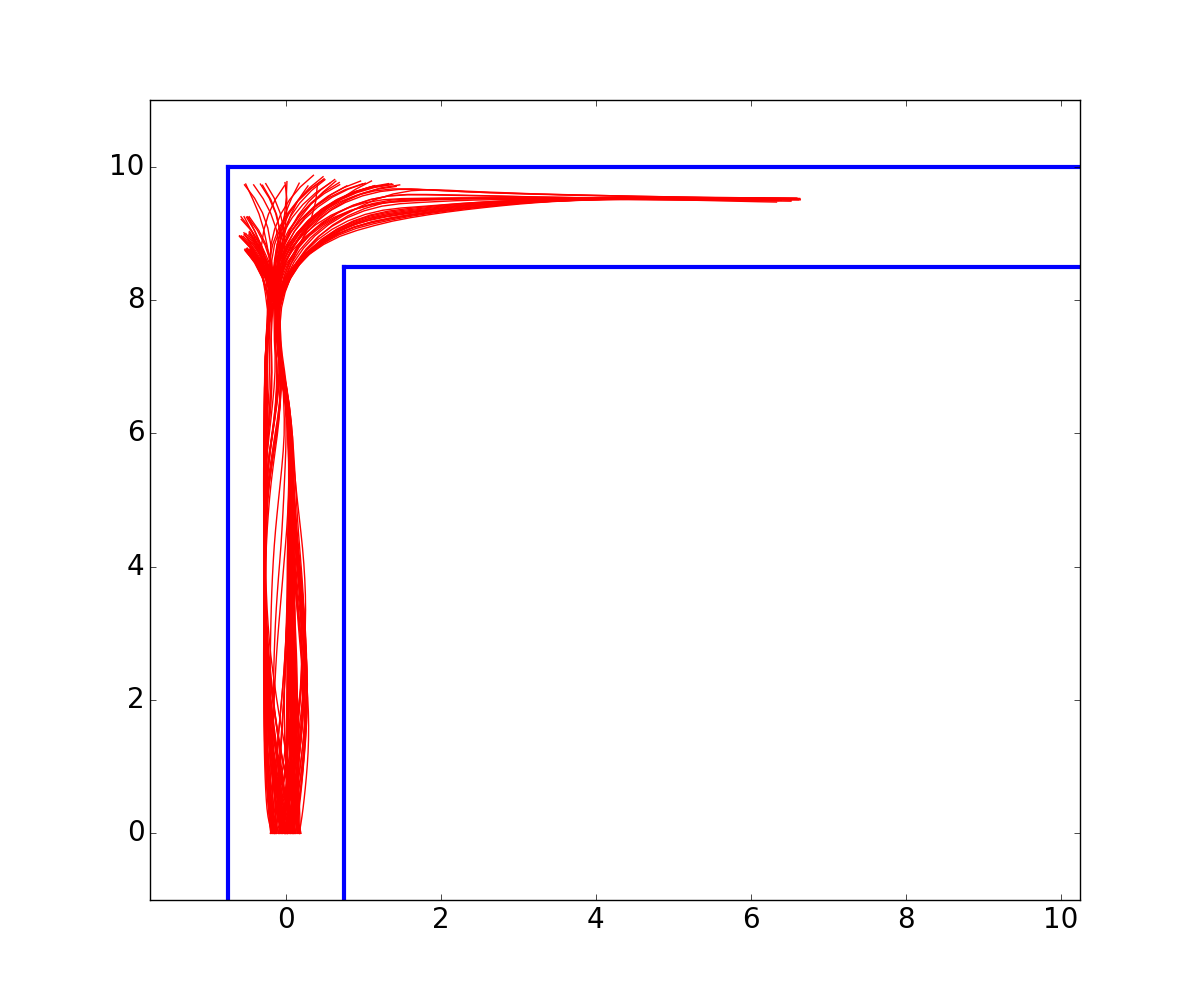}
           \label{fig:tanh64x64_DDPG1_faults}
       }
       \hfill
       \subfloat[DDPG, $128 \times 128$, controller 2: 51 \% safe.]{
           \includegraphics[width=0.23\linewidth]{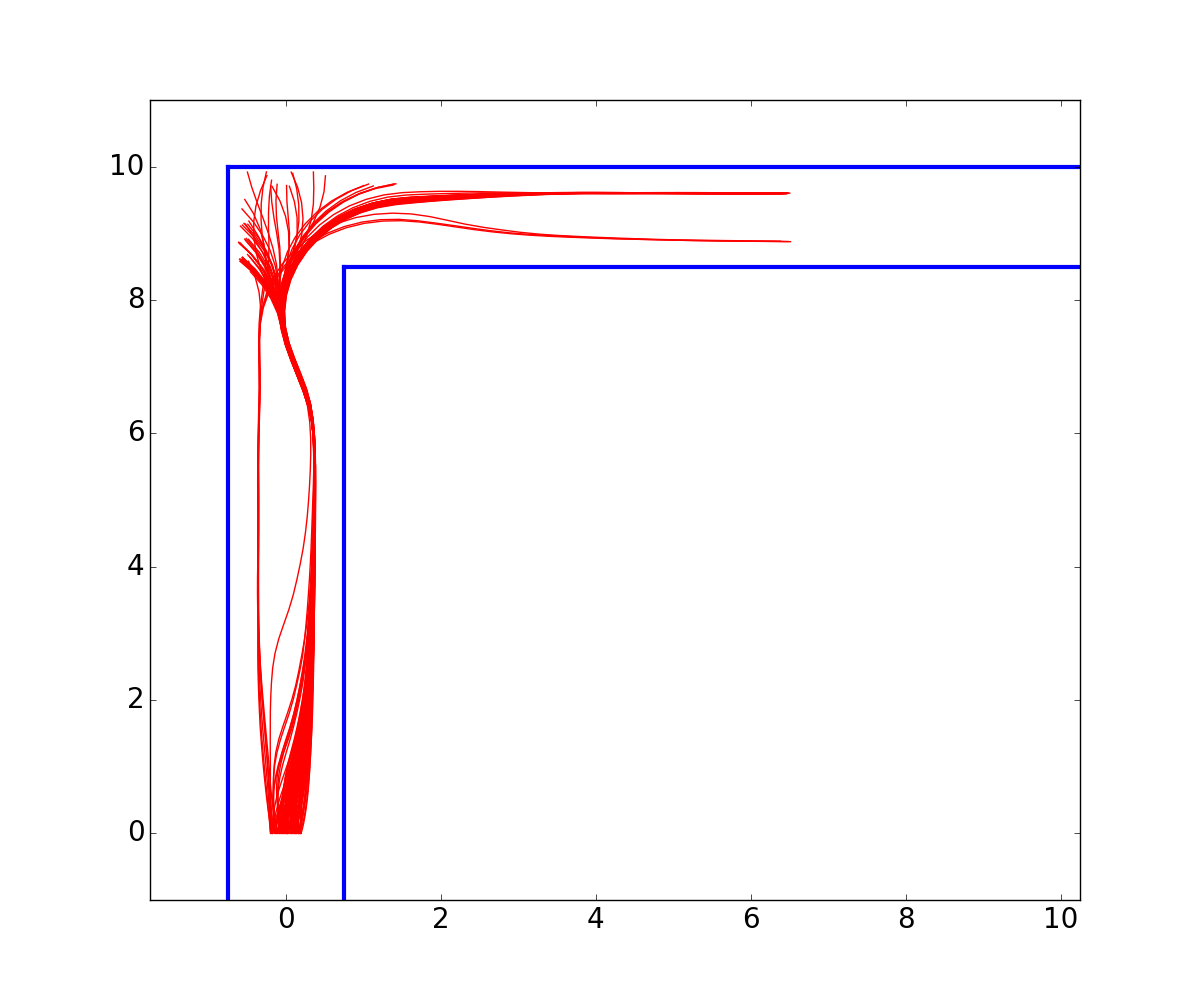}
           \label{fig:tanh128x128_DDPG2_faults}
       }
       \hfill
       \subfloat[TD3, $64 \times 64$, controller 1: 75 \% safe.]{
           \includegraphics[width=0.23\linewidth]{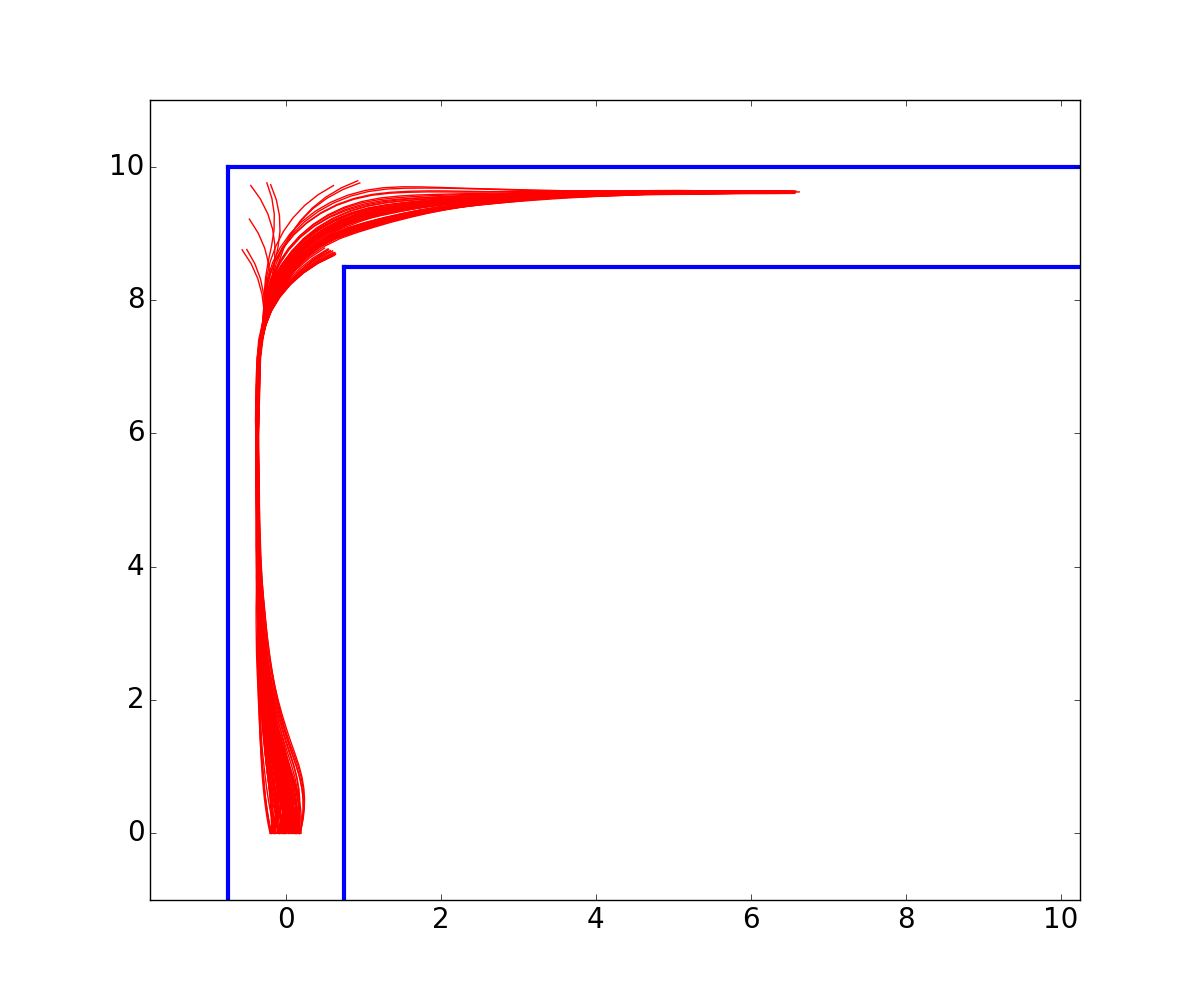}
           \label{fig:tanh64x64_TD3_1_faults}
       }
       \hfill
       \subfloat[TD3, $128 \times 128$, controller 1: 83 \% safe.]{
           \includegraphics[width=0.23\linewidth]{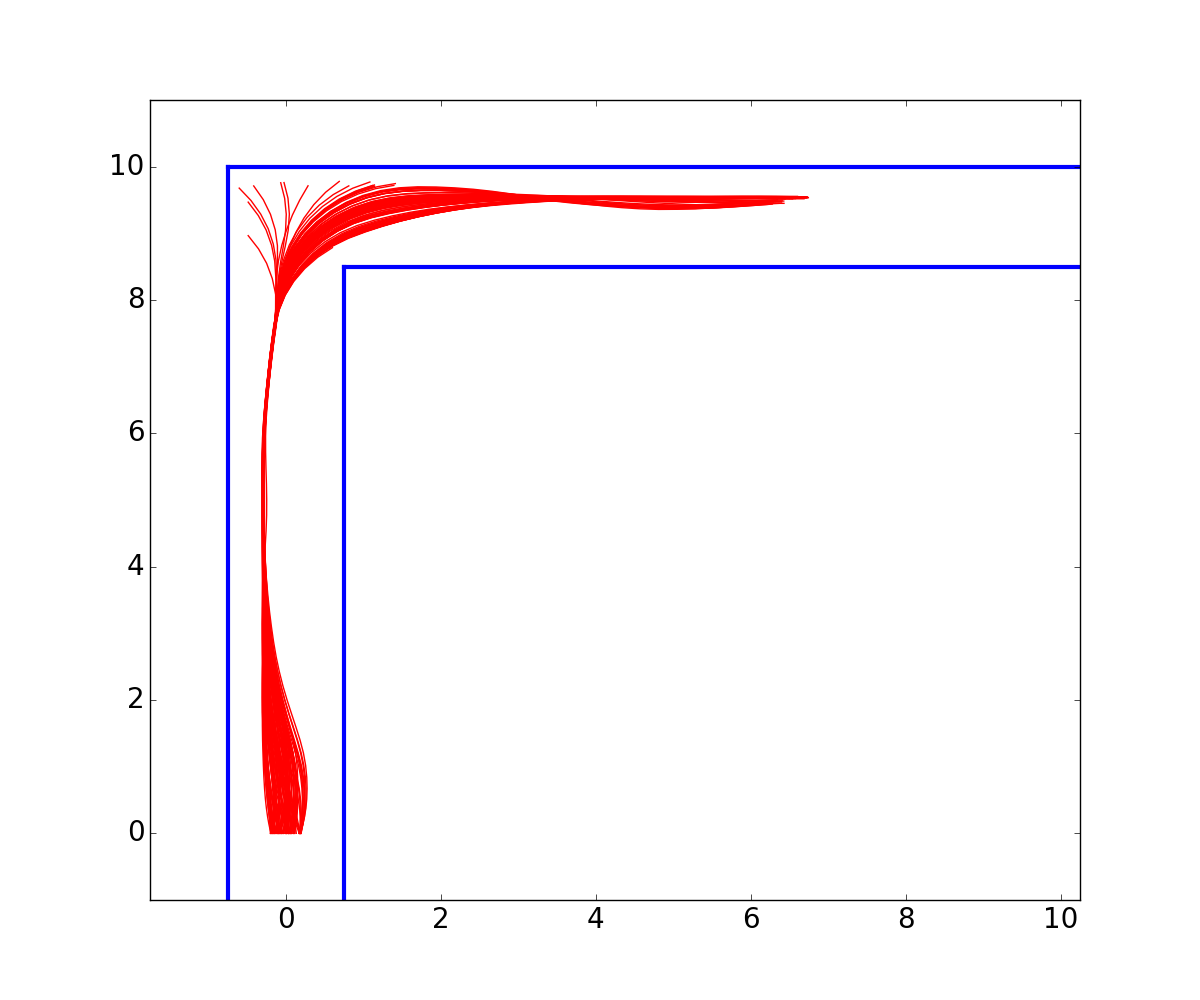}
           \label{fig:tanh128x128_TD3_1_faults}
       }
       \vspace{-10pt}
       \caption{Simulation traces for the NN controllers from
         Figure~\ref{fig:sim_traces}, with LiDAR faults added around
         the corner.}
       \label{fig:sim_traces_faults}
       \vspace{-10pt}
\end{figure*}

A second important observation is that the NN verification time is
roughly 10\% of the total verification time. This suggests that plant
verification, especially the observation model, is much more
challenging than NN verification. Thus, the scalability of
verification needs to be greatly improved not only in terms of the NN
size but also in terms of the plant complexity.

Finally, we note that the subset size is an indication of the
difficulty of verifying a given NN. The subset was decreased when the
uncertainty was so high that the safety property could not be verified
(some NNs could not be verified even for the smallest subset size
tried in our evaluation). Thus, a smaller subset size means that a NN
is potentially less robust to input perturbations. To illustrate this
point, we plot simulation traces for two NNs that either required
reducing the subset size or could not be verified at all and for two
NNs that were verified with the original subset size of 0.5cm. The
traces are shown in Figure~\ref{fig:sim_traces}. As can be seen, the
first two NNs are very sensitive to their inputs and produce
drastically different traces depending on the initial condition. As
shown in Section~\ref{sec:sim2real}, these NNs also result in unsafe
behavior in the real world.

\section{Exploring the sim2real Gap}
\label{sec:sim2real}
Having evaluated the scalability of current verification tools, we now
investigate the benefits and limitations of verification w.r.t the
real system. We perform experiments in an environment that is
identical to the verified one in terms of hallway dimensions, the main
difference being that the real environment contains reflective
surfaces that sometimes greatly affect LiDAR
measurements. Specifically, we note that the LiDAR model presented in
Section~\ref{sec:model} is fairly accurate when no reflections occur;
however, when a ray is reflected, it appears as if no obstacle exists
in that direction.

In order to assess the benefit of verification in an ideal
environment, we first cover most reflective surfaces and perform 10
runs per NN setup. All outcomes are reported in
Table~\ref{tab:sim2real}. As can be seen in the table, roughly 10\% of
runs in the modified environment were unsafe, uniformly spread across
different NNs, thus indicating that the LiDAR model is fairly accurate
when no reflections occur and that the verification result is strongly
correlated with safe performance. We emphasize that LiDAR faults
occurred even in this environment --
Figure~\ref{fig:scan_modified_environment} shows a LiDAR scan that
caused a crash.

Table~\ref{tab:sim2real} also shows that more crashes were observed in
the unmodified environment, due to multiple failing LiDAR rays (one
scan that led to a crash is shown in
Figure~\ref{fig:scan_unmodified_environment}). Interestingly, it is
possible to produce similar behavior in simulations as well --
Figure~\ref{fig:sim_traces_faults} shows the same runs as those in
Figure~\ref{fig:sim_traces}, but with five LiDAR rays randomly missing
around the area of the turn, similar to the pattern observed in
Figure~\ref{fig:scan_unmodified_environment}. The behavior illustrated
in Figure~\ref{fig:sim_traces_faults} is very similar to the real
outcomes reported in Table~\ref{tab:sim2real}, e.g., we observe
multiple crashes for setups DDPG $64 \times 64$, controller 1, and
DDPG $128 \times 128$, controller 2, while the TD3 NNs are more robust
to missing rays. However, although we can reproduce the LiDAR fault
model fairly well, training a NN that is robust to such faults is
challenging and was not possible with the DRL algorithms used in the
case study. Thus, it remains an open problem to train and verify a
robust NN for the problem considered in this paper.

\section{Discussion and Future Work}
\label{sec:discussion}
This paper presented a challenging verification case study in which an
autonomous racing car with a NN controller navigates a structured
environment using LiDAR measurements only. We evaluated a
state-of-the-art verification tool, Verisig, on this benchmark and
illustrated that current tools can handle only a small fraction of the
rays in a typical LiDAR scan. Furthermore, we performed real
experiments to assess the benefits of verification in terms of the
sim2real gap. Our findings suggest that numerous improvements are
necessary in order to address all issues raised by this case study.

\paragraph{Verification scalability in terms of the plant model}
As illustrated in the verification results in
Section~\ref{sec:verification}, the verification complexity scales
exponentially with the number of LiDAR rays. Thus, it is necessary to
develop a scalable approach that addresses this issue. For example,
one could use the structure of the environment in order to develop an
assume-guarantee approach such that verifying long traces may not be
required.

\paragraph{Verification scalability in terms of the NN}
Quantifying scalability in terms of the NN is not straightforward
since a large, but smooth, NN may be easier to verify than a small,
but sensitive, one, as indicated in Table~\ref{tab:verification}. Yet,
it is clear that existing tools need to scale beyond a few hundred
neurons in order to handle convolutional NNs, which are much more
effective in high-dimensional settings such as the one described in
this paper. While there exist tools that can verify properties about
convolutional NNs in isolation~\cite{wang18}, achieving such
scalability in closed-loop systems remains an open problem, partly due
to the complexity of the plant model as well.

\paragraph{Robustness of DRL}
Although DRL has seen great successes in the last few years, it is
still a challenge to train safe and robust controllers, especially in
high-dimensional problems. As shown in Section~\ref{sec:sim2real},
LiDAR faults can be reproduced fairly reliably in simulation; yet, we
could not train a robust controller using state-of-the-art learning
techniques. Thus, it is essential to develop methods that focus on
robustness and repeatability, with the final goal of being able to
verify the robustness of the resulting controllers.

\bibliographystyle{ACM-Reference-Format}
\bibliography{bibFile}

\end{document}